\documentclass[aps,prl,reprint,amsmath,amssymb,superscriptaddress,floatfix]{revtex4-2}

\usepackage{amsfonts}
\usepackage{xspace}
\usepackage{graphicx}
\usepackage{dcolumn}
\usepackage{bm}
\usepackage{units}
\usepackage{xcolor}
\begin{document}
\title{Enhancement of the Curie temperature in single crystalline ferromagnetic LaCrGe$_3$ by electron irradiation-induced disorder}

\author{E.~H.~Krenkel}
\affiliation{Ames National Laboratory, Ames, Iowa 50011, USA}
\affiliation{Department of Physics and Astronomy, Iowa State University, Ames,
Iowa 50011, USA}

\author{M.~A.~Tanatar}
\email[Corresponding author: ]{tanatar@ameslab.gov}
\affiliation{Ames National Laboratory, Ames, Iowa 50011, USA}
\affiliation{Department of Physics and Astronomy, Iowa State University, Ames,
Iowa 50011, USA}

\author{M.~Ko\'{n}czykowski}
\affiliation{Laboratoire des Solides Irradi\'{e}s, CEA/DRF/lRAMIS, \'{E}cole Polytechnique, CNRS, Institut Polytechnique de Paris, F-91128 Palaiseau, France}

\author{R.~Grasset}
\affiliation{Laboratoire des Solides Irradi\'{e}s, CEA/DRF/lRAMIS, \'{E}cole Polytechnique, CNRS, Institut Polytechnique de Paris, F-91128 Palaiseau, France}

\author{Lin-Lin Wang}
\affiliation{Ames National Laboratory, Ames, Iowa 50011, USA}

\author{S.~L.~Bud'ko}
\affiliation{Ames National Laboratory, Ames, Iowa 50011, USA}
\affiliation{Department of Physics and Astronomy, Iowa State University, Ames,
Iowa 50011, USA}

\author{P.~C.~Canfield}
\affiliation{Ames National Laboratory, Ames, Iowa 50011, USA}
\affiliation{Department of Physics and Astronomy, Iowa State University, Ames,
Iowa 50011, USA}

\author{R.~Prozorov}
\email[Corresponding author: ]{prozorov@ameslab.gov}
\affiliation{Ames National Laboratory, Ames, Iowa 50011, USA}
\affiliation{Department of Physics and Astronomy, Iowa State University, Ames,
Iowa 50011, USA}

\date{7 May 2024}

\begin{abstract}
LaCrGe$_3$  has attracted attention as a potential candidate for studies of quantum phase transitions in a ferromagnetic material. The application of pressure avoids a quantum critical point by developing a new magnetic phase. It was suggested that the disorder may provide an alternative route to a quantum critical point. We used low-temperature 2.5 MeV electron irradiation to induce 
relatively small amounts of point-like disorder in single crystals of LaCrGe$_3$. Irradiation leads to an increase of the resistivity at all temperatures with some deviation from the Matthiessen rule.  Hall effect measurements show that electron irradiation does not cause any detectable change in the carrier density. Unexpectedly, the Curie temperature, $T_{\text{FM}}$, \emph{increases} with the increase of disorder from approximately 90~K in pristine samples up to nearly 100~K in the heavily irradiated sample, with a tendency towards saturation at higher doses. Although the mechanism of this effect is not entirely clear, we conclude that it cannot be caused by effective ``doping" or ``pressure" due to electron irradiation. We suggest that disorder-induced broadening of a sharp peak in the density of states, $D(E)$, situated at $E_p=E_F-0.25$ eV below the Fermi energy, $E_F$, causes an increase in $D(E_F)$, leading to an enhancement of $T_\text{FM}$ in this itinerant ferromagnet. 
\end{abstract}

\maketitle

\section{Introduction}

Monotonic suppression of the transition temperature of second-order phase transitions to $T=0$ by a non-thermal tuning parameter (pressure, composition, disorder, uniaxial strain, magnetic field) provides a fruitful playground for the realization of novel exotic phases of matter \cite{QCP,PCCSLB2016}. For example, quantum fluctuations of the antiferromagnetic order parameter on approaching a quantum critical point (QCP) lead to unconventional superconductivity in heavy fermion, cuprate, organic, and iron-based superconductors \cite{Mathur,Louis,Matsuda}. Fluctuations of the charge density wave (CDW) order parameter lead to an increase in superconducting $T_c$ in (Ca,Sr)$_3$(Rh,Ir)$_4$Sn$_{13}$ compounds \cite{Klintengerg,Goh}. 

The application of similar ideas to ferromagnetic compounds has a fundamental problem. Quantum phase transitions in two-dimensional and three-dimensional metallic systems from a paramagnetic (PM) phase to a homogeneous ferromagnetic (FM) phase are generically first order, provided the material is sufficiently clean \cite{absenceFMQCP}. This is determined by the interaction of the magnetization with the electronic soft modes that exist in any metal, leading to a fluctuation-induced first-order transition.  The exceptions are non-centrosymmetric metals with strong spin-orbit coupling \cite{Kirkpatrick2020}.  As a result, when tuning material to QCP, the continuous FM-PM transition either becomes an abrupt first-order transition below a tricritical point or modulated magnetic phases appear; see Refs.\cite{Belitz2005,Dresden} for reviews.  However, it has also been shown that non-magnetic quenched disorder suppresses the tricritical temperature and the transition can remain second order down to zero temperature if the disorder strength exceeds a certain critical value, which opens new opportunities in studies of quantum criticality in small magnetic moment (fragile) ferromagnets \cite{PCCSLB2016,absenceFMQCP}. This suggests that disorder may play a pivotal role in quantum critical behavior of ferromagnets.

One of the potentially strong effects of point-like non-magnetic disorder is its impact on the energy distribution of the density of states (DOS), $D(E)$. The ferromagnetic instability in a metal occurs when the Stoner criterion $ID(E_F)>1$ is satisfied, here $I$ is the Stoner parameter of the strength of the Coulomb interaction. Normally, random point-like disorder leads to smearing of $D(E)$ causing a reduction of $D(E_F)$, thus destabilizing itinerant magnetism. However, if there is a peak in $D(E)$ not far from $E_F$, this smearing may cause an increase in $D(E_F)$ and enhance $T_\text{FM}$.

Substitutional disorder is an unavoidable component of systems in which composition variation is used as a tuning parameter, and disentangling its role from chemical doping/steric effects is very hard and sometimes impossible. Irradiation with energetic particles provides an alternative way to control the disorder. The use of disorder as a non-thermal tuning parameter, as, for example, in the cases of NbSe$_2$ \cite{NbSe2} and (Ba,K)Fe$_2$As$_2$ \cite{TimmonsBaK,irradiationdetwinning}, might provide a way to smear a first-order transition and enable a ferromagnetic quantum critical point \cite{Belitz2005,Dresden}. 

LaCrGe$_3$ is a typical example of avoided quantum criticality in a ferromagnetic system. Upon the application of hydrostatic pressure, the Curie temperature ($T_\text{FM}$ ) is initially suppressed, but then the nature of the transition changes to the first order and a new magnetic phase is stabilized \cite{Gati2021,32VT2016}.  When the magnetic transition becomes first-order at the tricritical point, application of a magnetic field $H$ along the magnetization axis reveals a wing structure phase diagram in the $T-p-H$ space, indicating the possibility of a new type of field-induced quantum critical point \cite{33Udhara2017}.   

Motivated by the idea of using disorder as a tuning parameter to reveal quantum criticality in itinerant metallic ferromagnets, we present a study the effects of a controlled disorder on magnetic ordering in LaCrGe$_3$. Disorder was introduced by low temperature electron irradiation, creating defects in form of vacancies with densities on the order of 10$^{-3}$ defects per atom   (i.e. comparable to the disorder associated with one out of 200 unit cells having a defect).  Contrary to theoretical predictions and general intuition, we found that the transition temperature to a ferromagnetic state in LaCrGe$_3$ is enhanced by the disorder.  The effect is most likely caused by the specifics of the electronic band structure of this material, which shows a strong peak in the density of states below $E_F$. 

\section{Methods} 

\subsubsection{Sample preparation}

Unlike initial growths of this material \cite{XiaoLin2013}, the single crystals of LaCrGe$_3$ used in this study were grown in two steps \cite{Mingyu2023} from melts of La$_{18}$Cr$_{12}$Ge$_{70}$ \cite{PCCNewmaterials,TSPCC} using fritted Canfield Crucible Sets (CCS) \cite{CCS,LSP}. First La$_{18}$Cr$_{12}$Ge$_{70}$ was heated  to 1150 $^\circ$C and cooled to 950 $^\circ$C over 50–100 h. At 950 $^\circ$C, a mixture of LaGe$_{2-x}$ plates and LaCrGe$_3$ rods was separated from the remaining liquid phase. Subsequently, this decanted liquid was resealed, heated to 1000 $^\circ$C to fully re-melt it, and then slowly cooled from 950 $^\circ$C to 825 $^\circ$C over roughly 100 h. At 825 $^\circ$C the growth was decanted and the resulting single phase LaCrGe$_3$ crystalline solid phase was separated from excess liquid.

Samples from the same batches used in our electron irradiation study were characterized by x-ray diffraction, resistance, and magnetization. The results were consistent with previous reports \cite{XiaoLin2013,32VT2016,33Udhara2017} in terms of the ferromagnetic Curie temperature ($T_\text{FM}$) and residual resistivity ratio (RRR), see Fig.~\ref{resistivitysamples} below. 

The rod-shaped single crystals with hexagonal cross sections were characterized by electrical resistivity and Hall effect measurements, with the electrical current along the long dimension of the sample corresponding to the $c-$axis. Contacts to the samples were soldered in a standard four-probe resistivity measurement configuration and 5-probe Hall effect configuration with indium. Conducting Dupont 4929N silver paste was used to mechanically stabilize the contacts \cite{FeSe}. The typical contact resistance was in the 10 to 100 m$\Omega$ range. Temperature-dependent resistivity and Hall effect measurements were performed in a {\it Quantum Design} Physical Property Measurement System (PPMS). Magnetic field for Hall effect measurements was applied along the $a-$axis, transverse to the current flowing along the $c-$axis. 
Resistivity measurements at 300~K were performed on 11 samples of LaCrGe$_3$ with current along the $c-$axis.  Two samples were discarded as obvious outliers, and from the remainder, the resistivity at room temperature was determined as 141 $\pm$ 12 $\mu \Omega \cdot$cm. In the remainder of this paper, we use this value for all pristine $c-$axis transport samples, and normalize geometric factors to match this value.  For the irradiation study, we selected four samples, shown in Fig.\ref{resistivitysamples}, with highly reproducible temperature-dependent resistivity curves that lay on top of each other and are indistinguishable to the eye. 

\begin{figure}[hbt!]
\includegraphics[width=\linewidth]{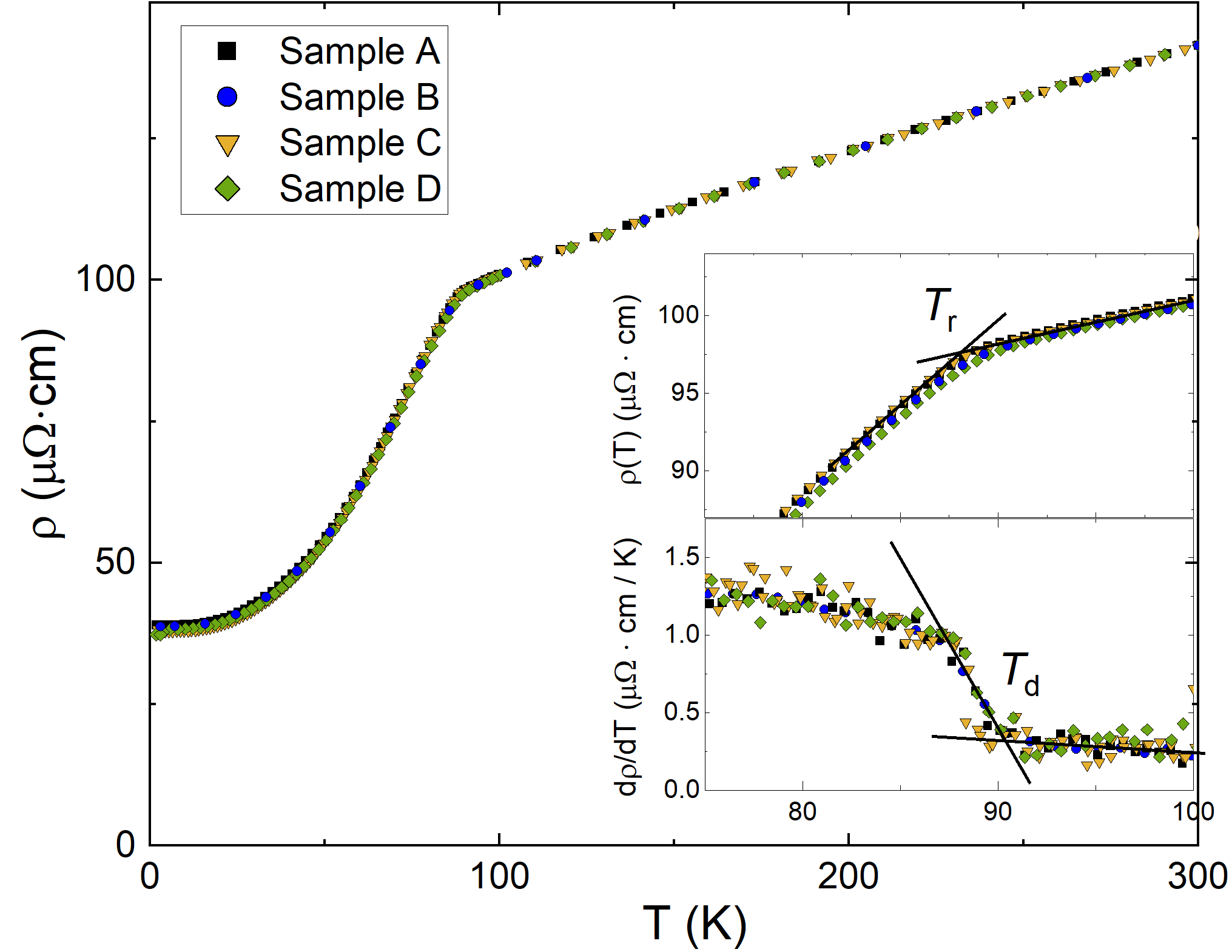}
\caption{
\label{resistivitysamples} 
Temperature dependent resistivity of the four selected samples of LaCrGe$_3$. The room-temperature resistivity of LaCrGe$_3$ for the current along $c-$axis was normalized to $141 \mu \Omega \cdot $cm, a value obtained from an array of 11 samples.  The geometric factors of all samples were adjusted to match this value. Insets zoom in on the transition temperature in resistivity (top inset), and temperature derivative of the resistivity (bottom inset). The two criteria used for the determination of the transition temperatures from resistivity data, $T_r$, and from the onset of the derivative increase, $T_d$, are shown.}
\end{figure} 

\subsubsection{Electron irradiation}

\begin{center}
\begin{table*}
\caption{Sample irradiation summary }
\begin{tabular}{| c|c|c|c|c|c|c|c |l|} 
\hline
Sample & Thickness & Dose & Entry & Exit E & Exit $\sigma$ & Exit & Attenuation  & $\rho (300\:K),( \% change)$ \\
  & ($\mu$ m) & $\; \text{C}/\text{cm}^2$ & $ \;\text{dpa} \times 10^{-4}$ & MeV & barn &  $\;\text{dpa} \times 10^{-4}$ & \%  &($\mu \Omega cm $\\ 
\hline 
\hline
A & 133 & 3 & 21.3 & 2.27 & 105 & 19.7 & 92 &150.0 (6\%)\\ 
\hline
B & 511 & 3 & 21.3 & 1.9 & 88 & 16.5 & 77  &149 (5.6\%)\\ 
\hline
C & 222 & 0.5 & 3.55 & 2.17 & 101 & 3.15 & 89  &142.0 (0.3\%)\\ 
\hline
D & 222 & 1 & 7.1 & 2.17 & 101 & 6.30 & 89  &150.5 (7\%) \\ 
\hline
\end{tabular}
\label{table}
\end{table*}
\end{center}

Electron irradiation was performed at the ``SIRIUS" Pelletron linear accelerator operated by Laboratoire des Solides Irradi\'{e}s at \'{E}cole Polytechnique in Palaiseau, France \cite{SIRIUS}. The 2.5 MeV electrons impact the sample kept in liquid hydrogen at 22 K to ensure efficient heat removal, and to prevent clustering and recombination of created atomic defects, vacancy-interstitial, called Frenkel pairs.
Due to their small rest mass, the relativistic energy transfer upon impact from electrons matches the energy of ionic displacement barriers. Heavier particles and ions induce correlated disorder from secondary collisions and cascades, such as extended clusters and columnar defects. The typical threshold energy to displace the ions is between 10 and 100 eV \cite{Damask1963,Thompson1969,annealing,Kyuilreview}. 

The interstitials usually have lower diffusion barriers compared to  vacancies. When the sample is returned to room temperature, some of the Frenkel pairs recombine, and some interstitials leave the crystal through various sinks, such as extended defects (dislocations/disclinations) and surfaces. The remaining metastable population of vacancies and interstitials is the point like disorder in our study \cite{Damask1963,Thompson1969}.  The exact rate of annealing is material-dependent and we use resistivity increase to quantify the added disorder.

The samples for resistivity measurements during and after electron irradiation were mounted on a thin mica plate, which was placed inside a hollow Kyocera chip C-QFN (Ceramic Quad Flat Non-Leaded Packages) \cite{Kyocera}. The Kyocera chip, mounted on a special holder, was inserted into the irradiation chamber and kept in liquid hydrogen during irradiation. 

For the analysis of the defect density we used an average cross-section for LaCrGe$_3$ at 2.5 MeV, $\sigma=114\;\text{barn}$. This value gives $7.1\times10^{-4}$ defects per atom per 1 C/cm$^2$. While the incoming electron beam energy is at 2.5 MeV, the energy is gradually lost and in a sample of LaCrGe$_3$ it should go to zero at the thickness of 1.3 mm (stopping distance). This limits the practical acceptable thicknesses of the samples to 0.5 mm or less. In Table ~\ref{table} we summarize parameters for the first irradiation run involving all the samples used in this study. For the thickest sample \#B the energy of the beam at the sample exit is calculated as 1.9 MeV. This energy is still sufficient to knock out all species of ions in the lattice \cite{Damask1963,Thompson1969}. Direct evidence for this conclusion comes from the identical increase of $T_\text{FM}$ and resistivity shift in samples \#A (0.133 mm thick) and \#B (0.511 mm thick) after receiving an identical dose of irradiation; see Figures \ref{resistivityirradiation}, \ref{resistivityirradiationHall} and Fig. \ref{shift} below. 

The flux of electrons was estimated by measuring the total electrical current through a 5 mm diameter diaphragm using a Faraday cup positioned behind the sample, so that only transmitted electrons were counted.  The dose of electron irradiation received by a sample is reported in $\text{C}/\text{cm}^2$. In conventional units, $1~\text{C}/\text{cm}^2=6.24\times10^{18}~\text{electrons}/\text{cm}^{2}$. Typically, the dose accumulated in a session varies from 0.5 (overnight) to 3 $\text{C}/\text{cm}^2$ (weekend). Larger doses are accumulated in several irradiation sessions, which can be separated by 6 to 12 months with ex situ sample characterization in between. Since the density of defects produced by irradiation is metastable, some annealing occurs at room temperature between runs. Throughout the manuscript we use ``pristine” and ``unirradiated” interchangeably to describe samples that have not been exposed to electron irradiation.   The doses shown in figures involving sample B are cumulative - 3 C /cm$^2$ from the first irradiation is listed as 3 C/cm$^2$.  When the same sample receives a second irradiation of 3 C / cm$^2$, it is shown in the figure as 6 C/cm$^2$ (for the total dose), and so on.  The initial doses for all samples are listed in column 2 of Table \ref{table}.

\subsubsection{Electronic band structure calculations}

Density functional theory (DFT) with local density approximation (LDA) as exchange-correlation functional has been used to relax structures and calculate the density of states (DOS) for LaCrGe$_3$ without and with vacancies. The DFT calculations have been done in Vienna Ab-initio Simulation Package (VASP) using projected augmented wave method and a plane-wave basis set with a kinetic energy cutoff of 227.2 eV. The primitive unit cell is fully relaxed on a (7$\times$7$\times$8) Monkhorst-Pack k-point mesh including the $\Gamma$ point with a Gaussian smearing of 0.05 eV and an increased kinetic energy cutoff of 1.25 times. The calculated lattice constants of 6.088 and 5.591 \AA ~agree with the previously calculated 6.078 and 5.587 \AA \cite{Nguyen}, which are slightly underestimated when compared to the experimental data of 6.165 and 5.748 \AA. For the accurate DOS calculation, a much denser (12$\times$12$\times$18) k-point mesh with tetrahedron method is used. The (3$\times$3$\times$3) supercell with vacancies is relaxed on a (2$\times$2$\times$3) k-point mesh and the corresponding DOS calculated on a denser (4$\times$4$\times$6) k-mesh. The absolute forces on ions are reduced to below 0.02 eV/\AA ~during relaxation. Previous DFT calculations have shown that LDA with the fully relaxed crystal structure can give a total magnetic moment of 1.09-1.12 $\mu_B$ in a good agreement to the experimental data of 1.22-1.25 $\mu_B$ \cite{XiaoLin2013,Mingyu2023,Nguyen}. Our calculated 1.14 $\mu_B$ with LDA also agrees with these results.

\section{Results} 

We begin by comparing the impact of irradiation on different samples. Figure~\ref{resistivitysamples} shows the temperature-dependent resistivity of four crystals of LaCrGe$_3$. The inset zooms on the area close to $T_\text{FM}$ in resistivity (top) and the temperature-dependent resistivity derivative (bottom). The selected samples show good reproducibility of both $T_\text{FM}$ and RRR, as well as the overall temperature dependence of the resistivity. Insets also show two ways of the transition temperature determination. The crossing points of the linear fits of the $\rho (T)$ curves above and below the transition was used to determine $T_r$ (upper inset). Similar linear fits  of the derivative curves (lower inset) were used to define $T_d$.  

\begin{figure}[hbt!]
\includegraphics[width=\linewidth]{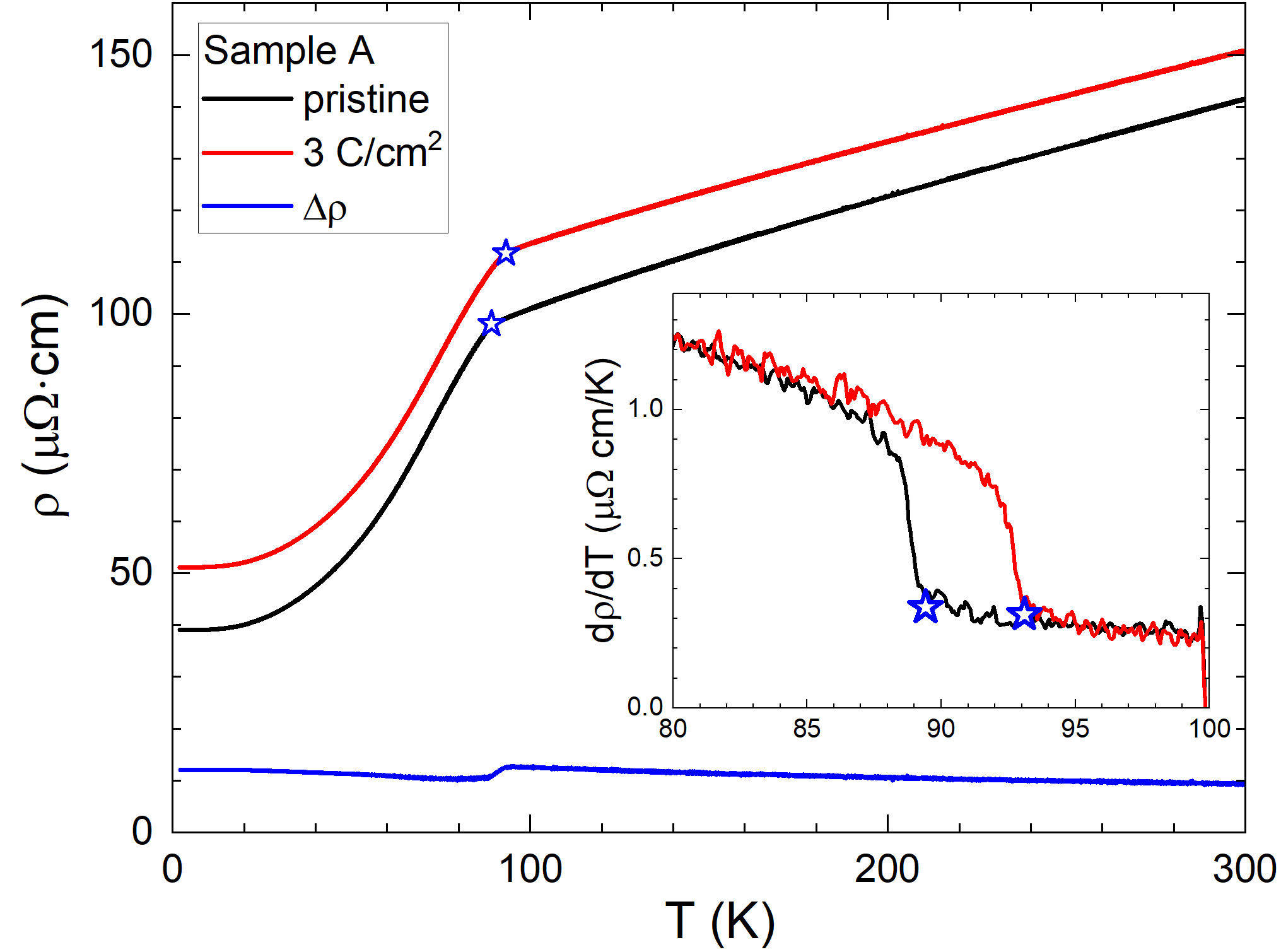}
\caption{
\label{resistivityirradiation} 
Temperature-dependent resistivity of sample \#A before and after 3 C/cm$^2$ of electron irradiation.  Stars mark the ferromagnetic transition temperature before and after irradiation using $T_r$ criterion. Inset shows the derivative of resistivity with respect to temperature, with stars defining the transition temperature using $T_d$ criterion. Blue line in the main panel shows the difference between resistivity curves before and after irradiation. }
\end{figure}

Figure~\ref{resistivityirradiation} shows the temperature-dependent resistivity of sample \#A before (black curve) and after 3 C/cm$^2$ of irradiation (red curve). The $\rho(T)$ curve demonstrates a nearly parallel shift with irradiation. The difference, $\Delta \rho(T)=\rho(T,3\:\text{C/cm}^2) - \rho(T, \text{pristine})$ (blue line, main panel), shows minor violation of Matthiessen's rule, increasing from  9.5 $\mu \Omega cm$ at 300 K to 12.5 $\mu \Omega cm$ at 95~K, just above $T_N$. Below the transition, $\Delta \rho$ drops to 10.5 $\mu \Omega cm$  and increases on further cooling to 10~K to 12 $\mu \Omega cm$, saturating at this value.   

Importantly,  $T_\text{FM}$ {\it increases} with irradiation, from 89.9 K to 92.9 K using $T_r$ criterion and from 90.6 K to 94.9 K using $T_d$ criterion, as shown in inset of  Fig.~\ref{resistivityirradiation}).  Because of the transition temperature increase, there is a notable feature in the difference plot at $T_\text{FM}$. 

\begin{figure}[hbt!]
\includegraphics[width=\linewidth]{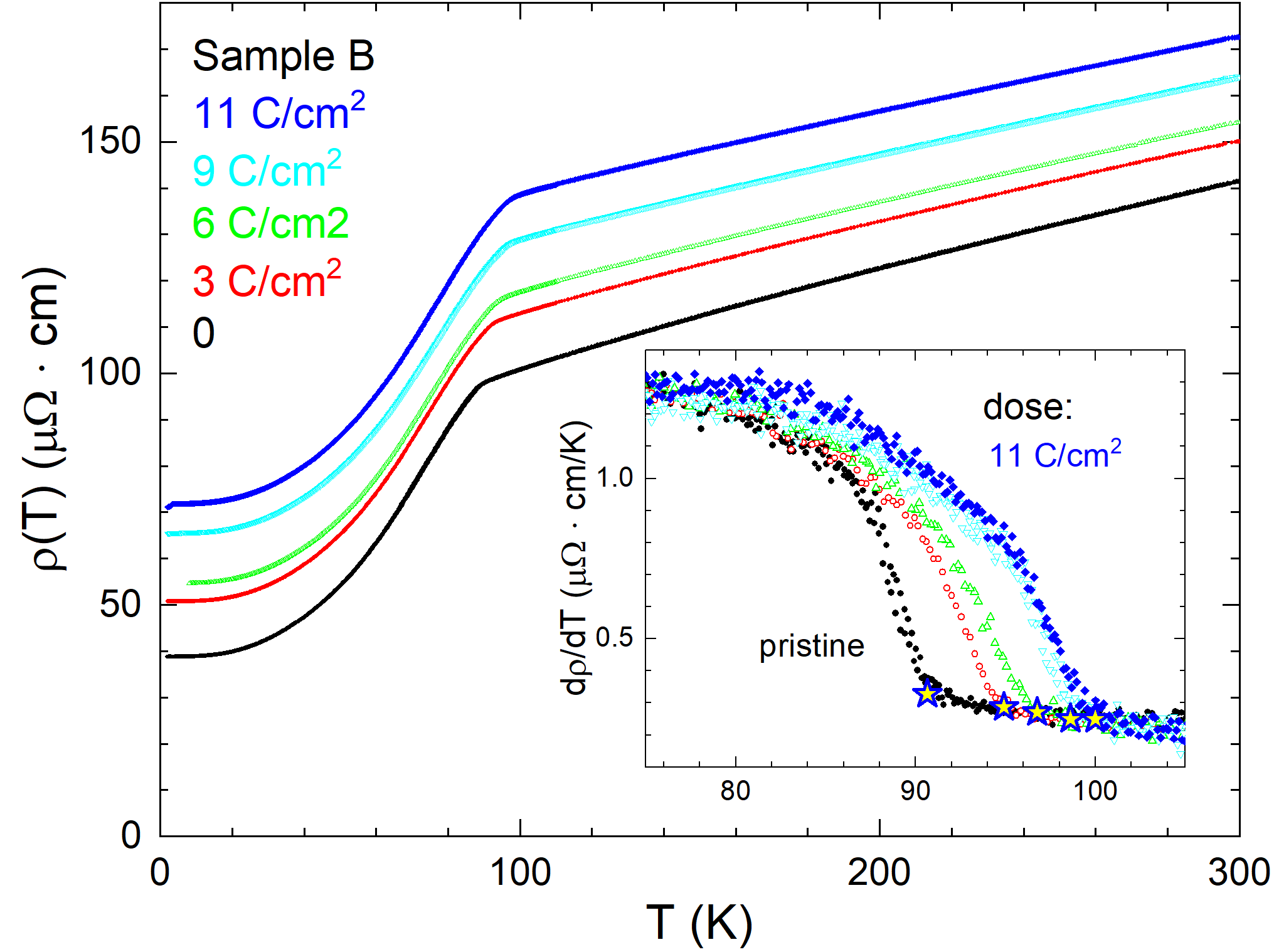}
\caption{
\label{resistivityirradiationHall} 
Evolution of the temperature dependent longitudinal resistivity of sample \#B with sequential electron irradiations. Inset shows the derivative of the resistivity with respect to temperature. Resistivity shows nearly parallel up-shift with irradiation dose. }
\end{figure}

\begin{figure}[hbt!]
\includegraphics[width=\linewidth]{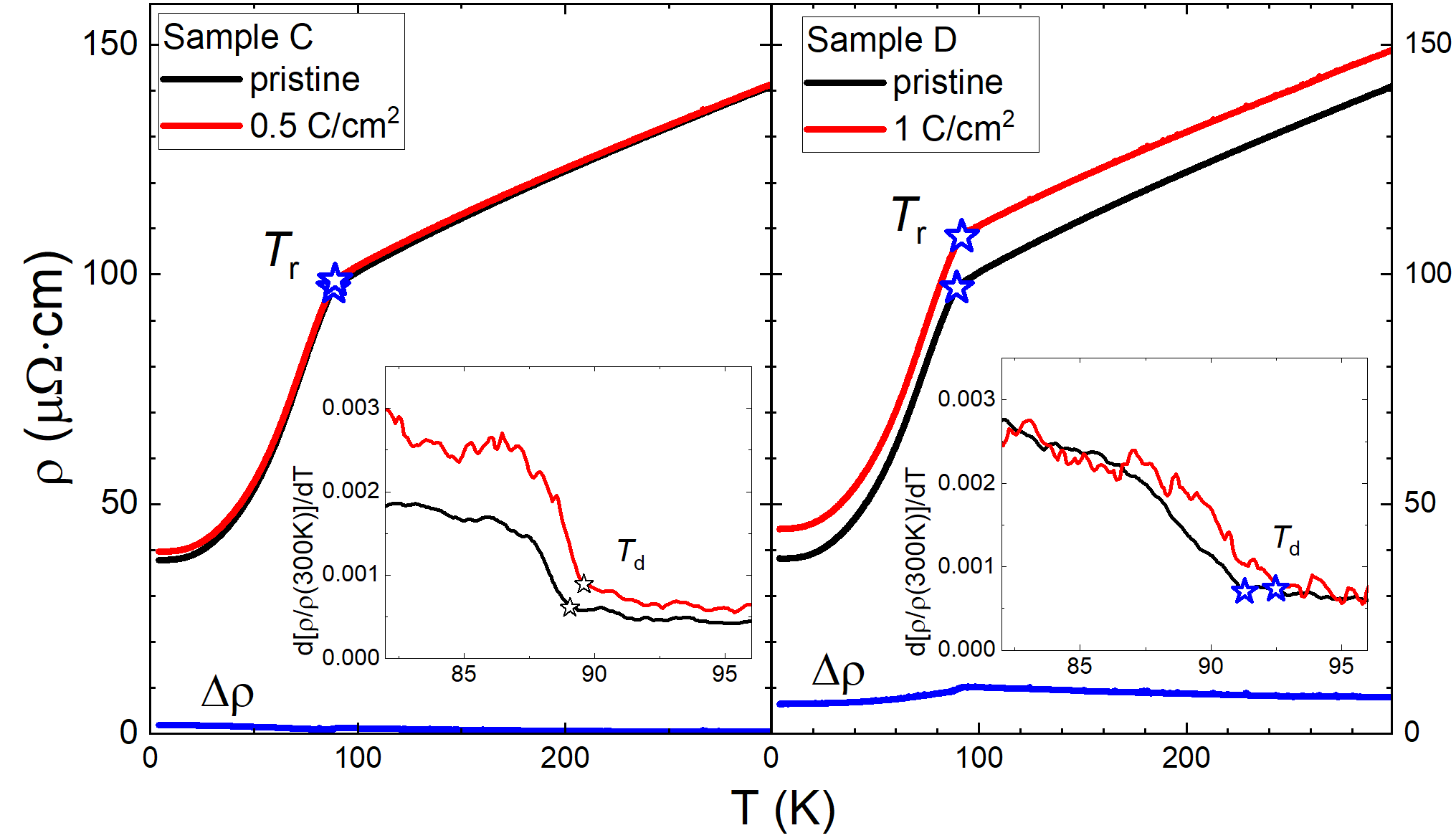}
\caption{Temperature-dependent resistivity of samples \#C (left panel) and \#D (right panel) before (black curves) and after (red curves) electron irradiation.  Stars mark the ferromagnetic transition temperature before and after irradiation using $T_r$ criterion.   Insets show the derivative of resistivity with respect to temperature, with stars defining the transition temperatures using $T_d$ criterion. Blue lines show the difference, $\Delta\rho=\rho(\text{irradiated})-\rho\text{(pristine)}$. }
\label{resistivityCD} 
\end{figure}

Figure~\ref{resistivityirradiationHall} shows the longitudinal resistivity of the sample \#B in zero magnetic field (also used for Hall measurements) in the pristine state (black curve) and after multiple irradiations with cumulative doses of 3~C/cm$^2$ (red), 6~C/cm$^2$ (green, second irradiation of 3C/cm$^2$ sample), 9 C/cm$^2$ (cyan, added 3 C/cm$^2$ to 6 C/cm$^2$) and 11 C/cm$^2$ (blue, added 2 C/cm$^2$ to 9 C/cm$^2$ sample). The irradiation for the 9 and 11 C/cm$^2$ doses was applied approximately 6 months after the previous doses,  while the 6 C/cm$^2$ dose was applied one year after the sample initially received 3 C/cm$^2$.  This explains the visibly smaller shift between the 3 and 6 C/cm$^2$ curves compared to the other doses, due to longer time for partial annealing of the sample at room temperature \cite{annealing,Kyuilreview}. Samples \#C and \#D were subjected to significantly smaller doses of irradiation (0.5 C/cm$^2$ and 1 C/cm$^2$ respectively) and are shown in Fig.~\ref{resistivityCD}.  Sample \#C followed the same trend in the transition temperature and resistivity with irradiation as samples \#A and \#B. Sample \#D showed a larger increase in both resistivity and the transition temperature per dose than the rest.

\begin{figure}[hbt!]
\includegraphics[width=\linewidth]{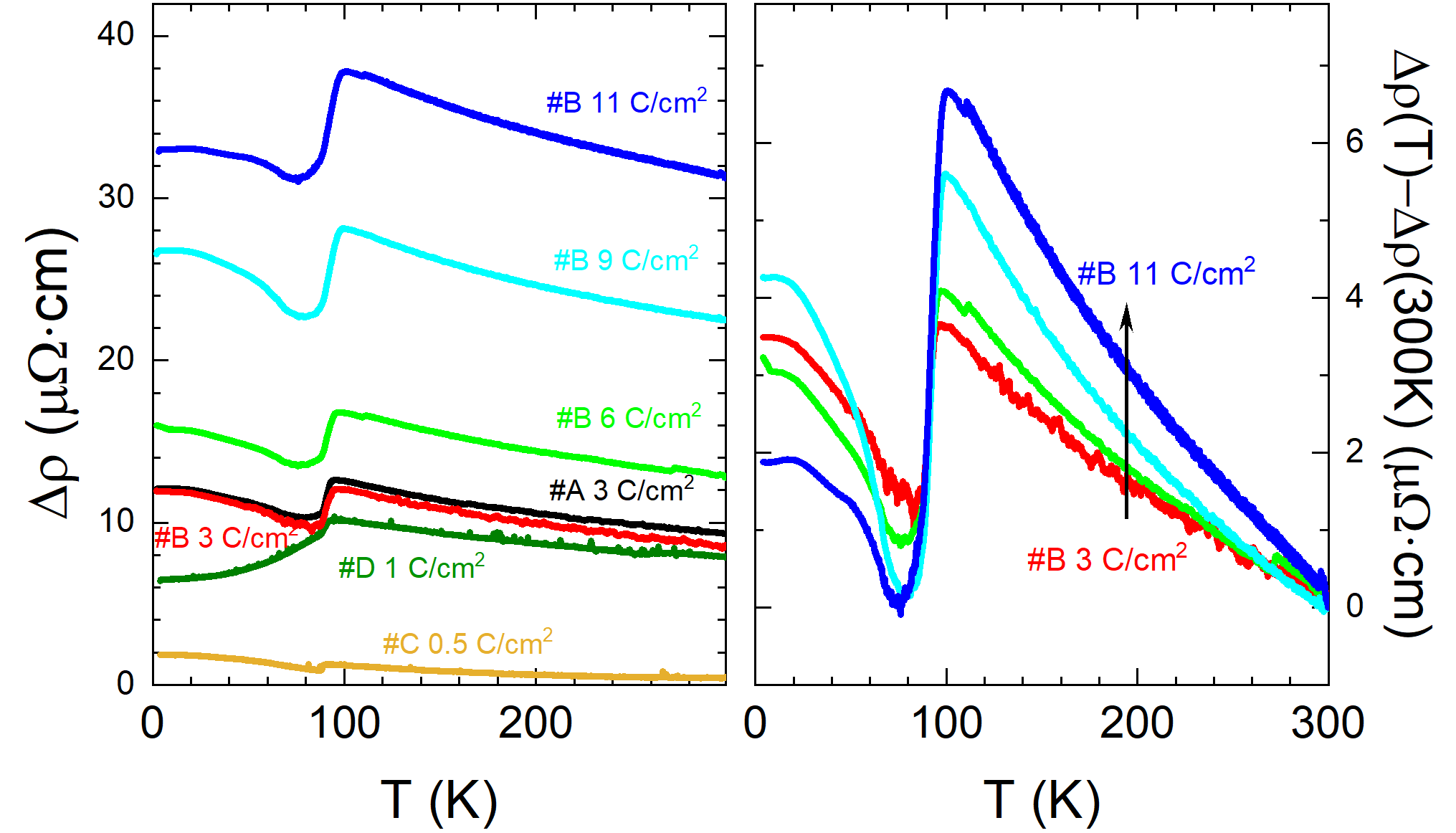}
\caption{
\label{shift}
Left panel:  Shift of the temperature dependent resistivity, $\Delta \rho(T,\text{Dose})=\rho(T,\text{Dose})-\rho(T,0 \text{C}/\text{cm}^2$). Sample \#A, after a dose of 3 C/cm$^2$ (black); sample \#B after multiple irradiations with cumulative doses of 3 C/cm$^2$ (red), 6 C/cm$^2$ (light green), 9 C/cm$^2$ (cyan) and 11 C/cm$^2$ (dark blue); sample \#C after 0.5 C/cm$^2$ (yellow); and sample \#D after 1 C/cm$^2$ (dark green) irradiation. The right panel shows the relative shifts in resistivity with respect to their values at room temperature, $\Delta \rho(T)- \Delta \rho (300 K)$.  }
\end{figure}

The temperature-dependent shift, $\Delta \rho$, for all samples is compared in Fig.~\ref{shift}. It has the same shape for samples \#A, \#B and \#C, showing slight increase on cooling except for in the transition area. 

\begin{figure}[hbt!]
\includegraphics[width=8cm]{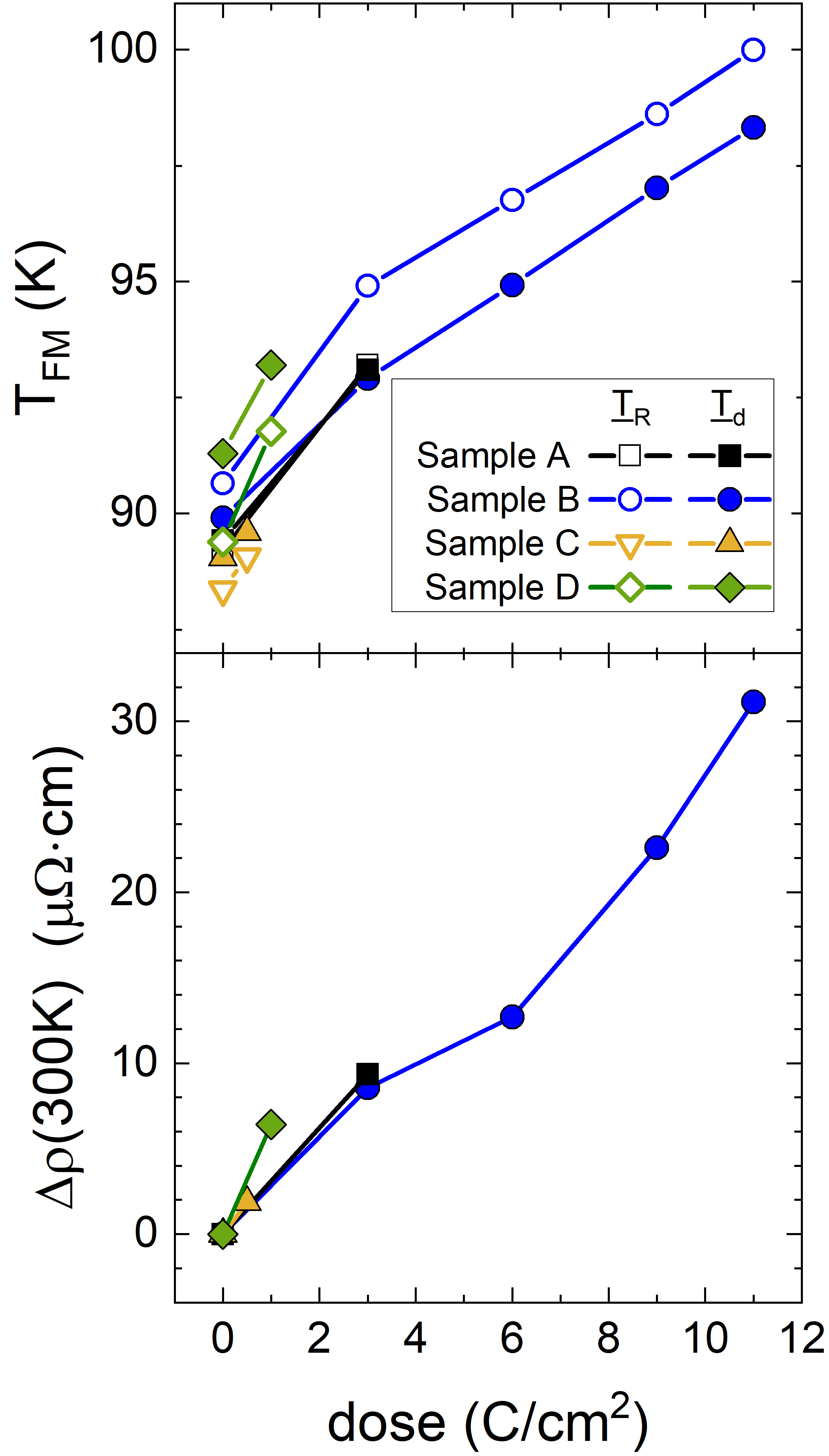}
\caption{
\label{dose}
 Top panel:  Dose dependence of ferromagnetic Curie temperature $T_\text{FM}$ determined using $T_r$ (closed symbols) and $T_d$ (open symbols) criteria. Bottom panel:  Dose dependence of the shift of the room temperature resistivity $\Delta \rho(300\:\text{K}, \text{Dose})=\rho(300\:\text{K},\text{Dose})-\rho(300\:\text{K},0)$.}
\end{figure} 

\begin{figure}[hbt!] 
\includegraphics[width=\linewidth]{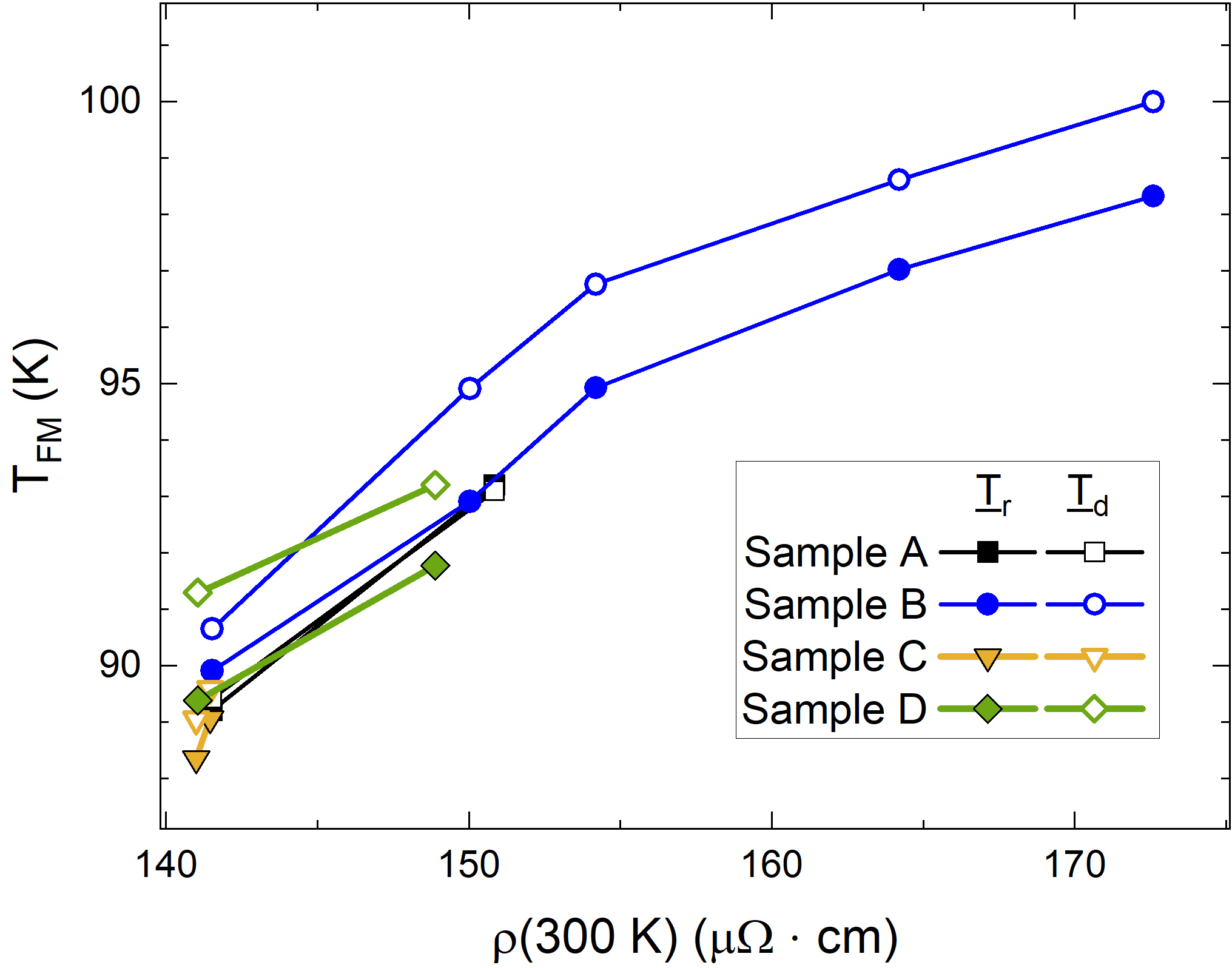}
\caption{
\label{TFMrho300K}
Change in the ferromagnetic transition temperature in single crystals of LaCrGe$_3$ as a function of the room temperature resistivity, which is used here as a proxy for the amount of disorder. } 
\end{figure}

The shift in the resistivity at 300~K is a reasonable proxy for defect concentration. The bottom panel of Fig.\ref{dose} presents variation of resistivity at $\rho(300\:\text{K})$ as a function of irradiation dose. The top panel presents dose dependence of the ferromagnetic   transition temperatures determined using $T_r$ (closed symbols) and $T_d$ (open symbols) as criteria. The dependence of the ferromagnetic transition temperature on $\rho(300\:\text{K})$  is shown in Fig.~\ref{TFMrho300K}.  It is clear, that  despite some variation of $T_\text{FM}$ in the pristine state, all samples show a consistent trend of $T_\text{FM}$ increase with increased disorder.  When sample \#B was subjected to higher doses of irradiation, it revealed a trend toward $T_\text{FM}(\rho)$ saturation.

\begin{figure}[hbt!]
\includegraphics[width=\linewidth]{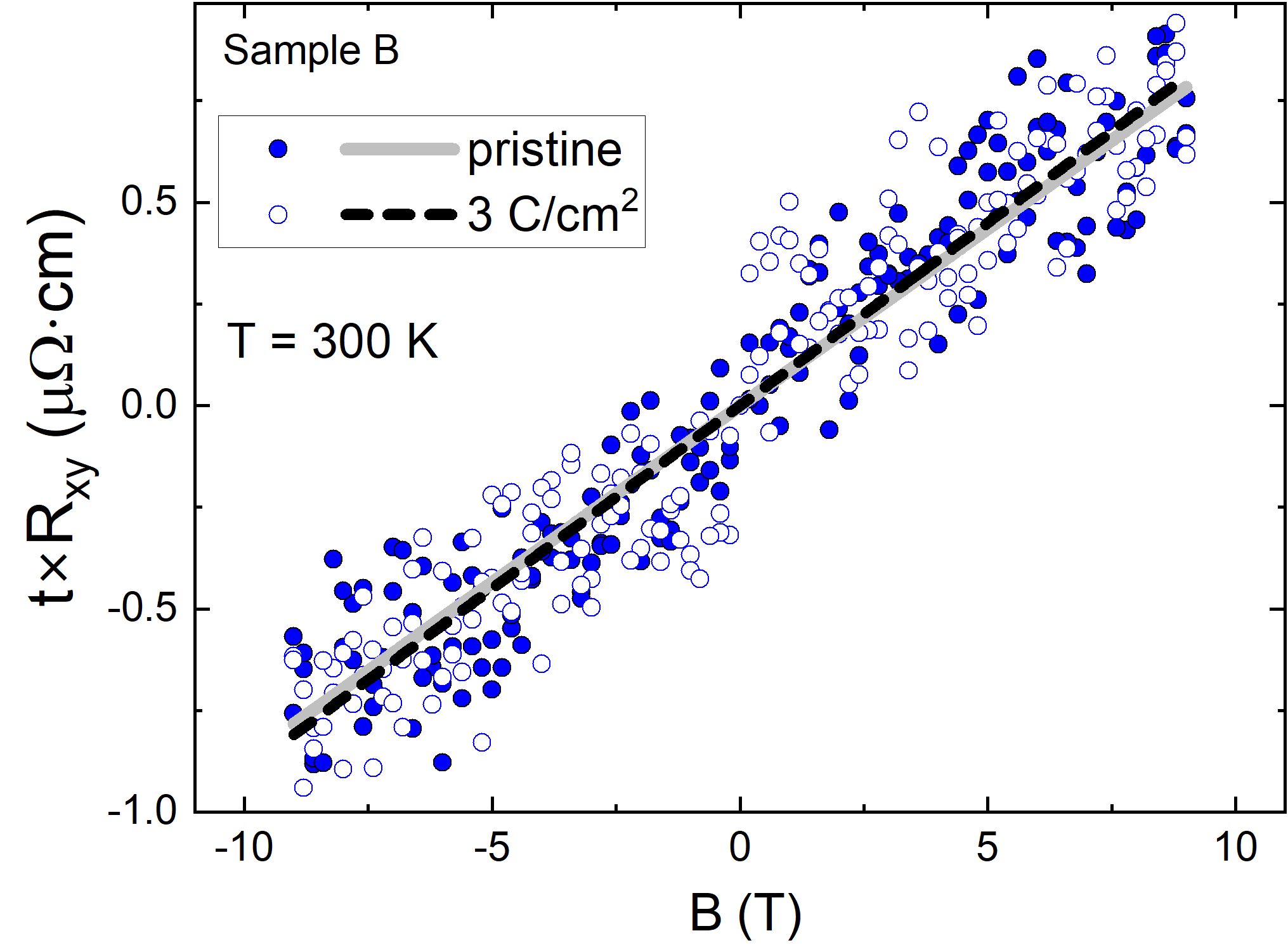}
\caption{
\label{Hall}
A product of the transverse Hall resistance $R_{xy}$ of sample \#B and its thickness $t$ before and after irradiation as a function of magnetic field.  Lines show the linear fit used for finding the Hall coefficient at 300~K. At this temperature, the difference in carrier density with irradiation is insignificant to within error ($9.0 \pm 0.2 \cdot 10^{-4}$ cm$^3$ / C vs. $8.7 \pm 0.2 \cdot 10^{-4}$ cm$^3$ / C), showing that irradiation does not dope the system.}
\end{figure}

Finally,  Fig.~\ref{Hall} presents field-dependent Hall resistance $R_{xy}$ multiplied by sample thickness in sample \#B at room temperature, in the paramagnetic state far above the long range ordering at $T_\text{FM}$. The data in the pristine state are shown with solid symbols, and after 3 C/cm$^2$ electron irradiation with open symbols. Unfortunately, the contact for the transverse (Hall) resistance measurement deteriorated after 6 C/cm$^2$ of irradiation and mechanically detached after 9 C/cm$^2$. For both data sets the Hall resistance is linear in magnetic field,  and have identical linear fit slopes (grey and dashed lines) to within error ($9.0\pm0.2\times10^{-4}$~cm$^3/C$ to $8.7\pm0.2\times10^{-4}$~cm$^3/C$). This suggests that the change (increase) of the carrier density is less than 3\%, if any. This should be compared to the 6\% change of resistivity for the same doses. The sign of the Hall effect is consistent with the sign of reported thermopower \cite{thermopower}.

\section{Discussion}

Summarizing our observations, we see three effects of controlled disorder on the properties of LaCrGe$_3$ crystals. (1) The resistivity increases with disorder roughly independently of temperature and reveals only minor deviations from the Matthiessen's rule both above and below $T_\text{FM}$.  These deviations diminish further with dose increase. (2) The resistivity increase is not accompanied by any sizable change of the Hall effect in the paramagnetic phase well above $T_\text{FM}$. This suggests that irradiation primarily changes the elastic scattering rate and is not changing the Fermi surface.  (3) The ferromagnetic transition temperature, T$_\text{FM}$, is increased nearly linearly as a function of dose for small amounts of irradiation in all samples studied.  At high doses there is a tendency toward saturation.  

\begin{figure}[hbt!]
\includegraphics[width=\linewidth]{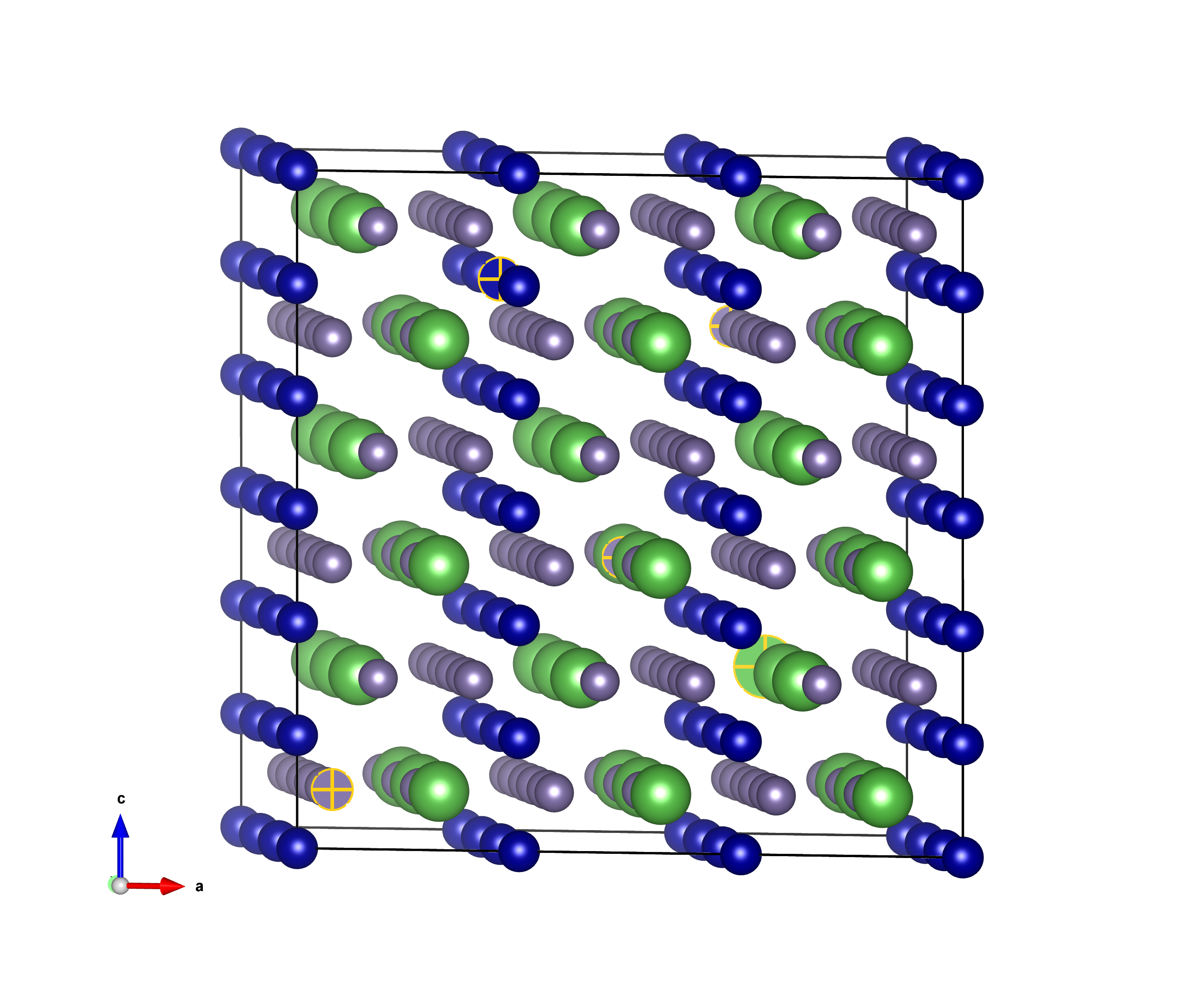}
\includegraphics[width=\linewidth]{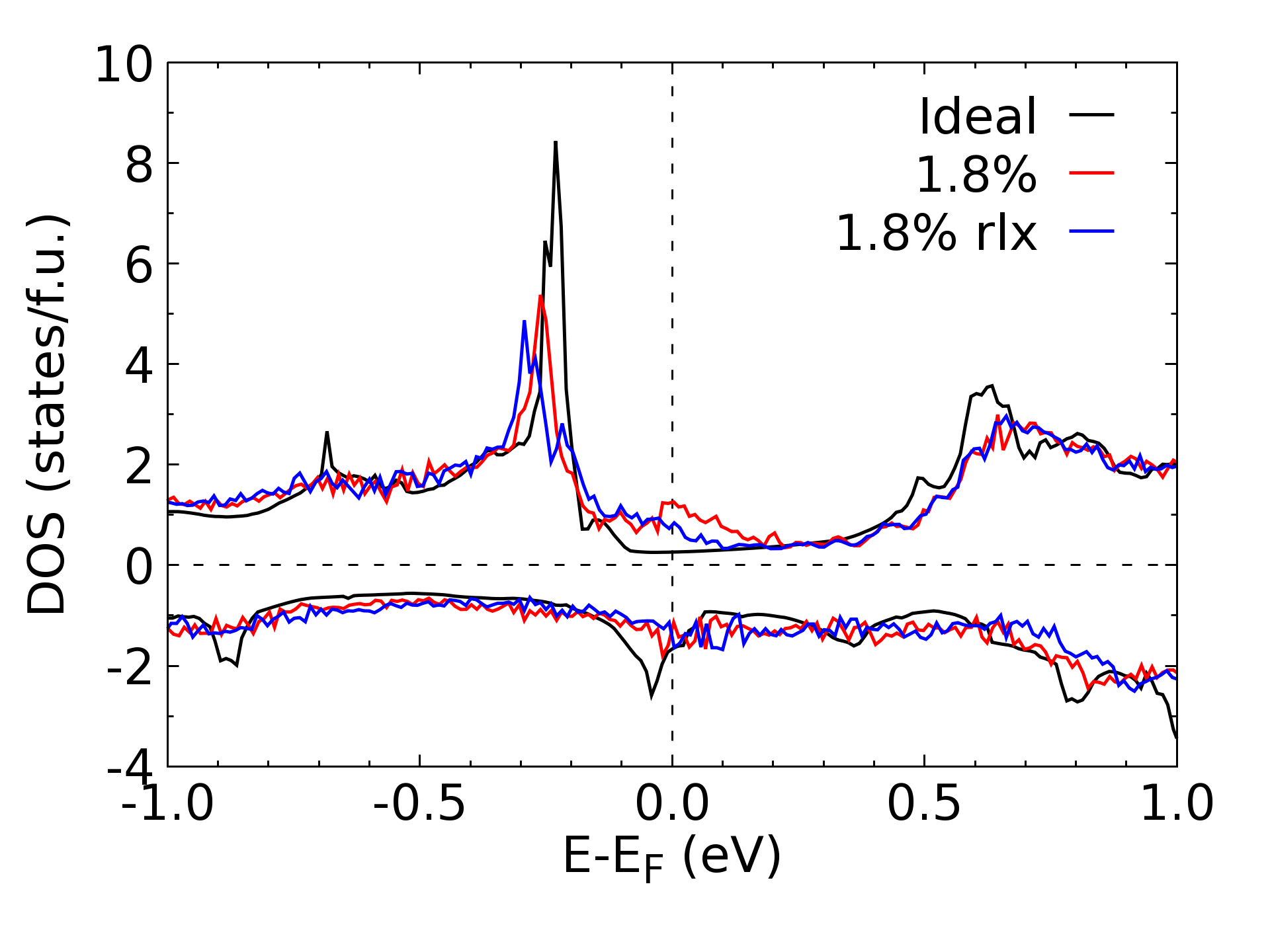}
\caption{Top panel: the supercell structure of LaCrGe$_3$ with La, Cr and Ge ions shown by green, blue and grey spheres, respectively. One formula unit (i.e. one La, one Cr and three Ge atoms labeled with yellow crosses) are removed to create vacancies of 1.8\% to represent disorder. Bottom panel: the density of states, $D(E)$, of ferromagnetic LaCrGe$_3$ in pristine state (black line) and in a (3$\times$3$\times$3) supercell with 1.8\% vacancies, as shown in the top panel. Red line is calculation with the rigid structure, blue line is calculation after ionic relaxation. The spin majority (minority) DOS is plotted as positive (negative) value. The sharp DOS peak in the pristine state near $E_F-0.25$~eV is broadened by vacancies and leads to a notable increase in the DOS at $E_F$. } 
\label{DOS}
\end{figure} 

Theoretically, for a general case of an itinerant system, an increase in disorder suppresses the ferromagnetic transition temperature, $T_\text{FM}$ due to the smearing of DOS leading to the reduction of $D(E_F)$. Is it possible that similar smearing could instead result in an increase of $D(E_F)$? In Fig.~\ref{DOS} we present DOS in LaCrGe$_3$ calculated for the pristine structure and for the structure with 1.8\% vacancies, the lowest concentration of vacancies suitable for a (3$\times$3$\times$3) supercell, as shown in the top panel. The average distance among these vacancy sites is about 8~\AA.  The vacancy sites are separated among themselves by at least a full unit cell in all directions. The characteristic feature of band structure of LaCrGe$_3$ is a sharp peak in $D(E)$ located near $E_F-0.25$~eV (black line) \cite{Nguyen} originating from the almost flat Cr 3$d$ bands ($d_{xz}$ and $d_{yz}$ orbitals) in the spin majority along the $\Gamma$-M-K-$\Gamma$ direction. Noticeably DOS right at $E_F$ for spin majority is very small and even smaller than that of the spin minority. Vacancies induce disorder by disrupting chemical bonds. As shown in the bottom panel of Fig.~\ref{DOS}, 1.8\% vacancies notably reduce and broaden the peak near $E_F-0.25$~eV (red line). However, as the consequence of the broadening, the DOS at $E_F$ shows a sizable increase for the majority spin, while it slightly decreases for the minority spin. Such opposite changes in the DOS at $E_F$ vs. $E_F-0.25$~eV are also true with ionic relaxation (blue line). The overall DOS at $E_F$ increases by 0.5 states/f.u. or 25\% accounting for both spin channels. The broadening of almost-flat bands at E$_F$-0.25 eV leads to an increase of the DOS at E$_F$. 

The calculated significant increase of DOS for a sample with 1.8\% vacancies, should be scaled down to match the significantly lower densities of the vacancies introduced by irradiation in our experiments. As we have shown, the actual density is about one order of magnitude smaller for the highest doses achieved. Although an increase in $D(E_F)$ provides a plausible mechanism for $T_c$ increase, it should lead to a slope change of resistivity curve due to carrier density variation, violation of Matthiessen's rule and a change in the Hall effect. Neither is obvious in our experiment, which may suggest that the states from the broadened flat bands remain localized.  

Suppression of the ordering transition temperature with disorder is a general trend for materials with different types of order.  Examples include a nematic/antiferromagnetic transition in the parent iron-based superconductor  BaFe$_2$As$_2$,  isoelectron substituted BaFe$_2$(As$_{1-x}$P$_x$)$_2$ \cite{Taka}, and in hole-doped (Ba$_{1-x}$K$_x$)Fe$_2$As$_2$ \cite{npj}. Similar suppression of the charge density wave transition temperature with disorder is found in NbSe$_2$ \cite{Mutka,NbSe2}, TaSe$_2$ and TaS$_2$ \cite{Mutka,Petrovic}.  Disorder suppresses the transition temperature in nodal superconductors \cite{nodal}, but the $T_c$ is practically insensitive to disorder in conventional $s-$wave superconductors (known as Anderson theorem) \cite{Anderson}, with the exceptions coming from the effect of strong disorder on the density of states as observed in aluminum \cite{Aluminum}. 

The exceptions to this common trend of ordering temperature suppression are mostly found in systems with competing orders. For example,  in NbSe$_2$ \cite{Mutka,NbSe2}, TaSe$_2$ \cite{Mutka}, TaS$_2$ \cite{Mutka,Petrovic}, Lu$_5$Ir$_4$Si$_{20}$ \cite{Ulrich1}, cuprates \cite{Ulrich2} and some Remeika 3-4-13 compounds \cite{Elizabeth} superconducting $T_c$ increases on suppression of CDW. In all these materials two electronic orders are competing for the density of states at the Fermi surface, and disorder tips the balance in favor of subdominant order (superconductivity) by suppressing dominant (higher transition temperature) order \cite{Gabovich}. Similar phenomena are observed in FeSe, in which superconductivity is competing with nematic order \cite{Serafim}. 
The only exception to this rule is observed in (Ba$_{1-x}$K$_x$)Fe$_2$As$_2$ where dominant $C_2$ stripe antiferromagnetic order is stabilized when the subdominant (lower transition temperature) $C_4$ phase is suppressed by disorder \cite{TimmonsBaK}.
The possibility of competing phases in LaCrGe$_3$ is suggested by the observation of several types of ordering in the pressure phase diagram \cite{Gati2021}. Another ferromagnetic phase was suggested to be stabilized in LaCrGe$_3$ at ambient pressure below $\sim$70~K \cite{Mingyu2023,Valentin2023}. 

Another mechanism that could potentially explain the observed change in $T_\text{FM}$  would be an effective change in the average volume of the unit cell due to induced atomic defects. Experiments on neutron-irradiated Cu \cite{n-irr-Cu}, electron-irradiated Cu\cite{irradiationstress} and electron-irradiated Al \cite{e-irr-Al} showed that the lattice expands at low temperature, which is equivalent to a ``negative effective pressure".  When measurements were performed as a function of temperature, it was found that at room temperature about 16\% of the effective lattice parameter change at low temperatures, $\Delta a/a$, remained in copper and no resolvable leftover strain was observed in aluminum. The annealing process has several stages and is quite complicated.  In this scenario, the effect of irradiation would result in an effective shrinkage of the lattice, exerting "positive effective pressure". Therefore, the sign of the effect depends on a material and conditions of irradiation and thermal history of the sample. 

Regardless the sign, let us estimate the volume change due to defects produced by irradiation, which will allow comparison with studies of LaCrGe$_3$ under pressure. One way is to estimate the total volume of all vacancies, assuming that no interstitials are left. Irradiation with the dose of 1 C/cm$^2$ induces about $3.5\times 10^{-3}$ vacancies per conventional unit cell, which is approximately 0.005 \AA$^3$. With a unit cell volume of 218.5 \AA$^3$, this gives $\Delta V/V \approx 2.2 \times 10^{-5}$ change. Our highest irradiation dose was 11 C/cm$^2$, see Fig.\ref{resistivityirradiationHall}, which results in $\Delta V/V \approx 2.5 \times 10^{-4}$ total volume change. On the opposite side of the estimate, we can assume that all defects stay and result in a rate of change reported for Cu and Al in Ref.\onlinecite{irradiationstress}. Remarkably, it was found that at low temperatures, the rate of change, $\eta=(\Delta V/V)/\Delta \rho$, does not vary much for a wide range of irradiation conditions \cite{n-irr-Cu,irradiationstress,e-irr-Al}. For example, for copper irradiated with 2.6 MeV electrons at 6 K (quite similar to the parameters of our work), $\eta=1.75\;(\mu\Omega\cdot\text{cm})^{-1}$ was reported \cite{irradiationstress}.  As shown in Fig.\ref{resistivityirradiationHall}, 11 C/cm$^2$ results in a resistivity change of $\Delta \rho\approx38\;\mu\Omega\cdot\text{cm}$. This gives $\Delta V/V \approx 4.4 \times 10^{-4}$ total volume change, which is of the same order as the first estimate. However, the reported change of resistivity in the LaCrGe$_3$ was measured after the irradiated sample was brought up to the room temperature. According to Ref. \cite{irradiationstress} only about 16\% of the effect remains for the linear change of the lattice parameter $\Delta a/a$. Therefore, the estimate becomes $\Delta V/V \approx 3.5 \times 10^{-6}$ for the total volume change, which is much smaller than the first estimate.

In our experiments, the enhancement of $T_\text{FM}$ per 1 C/cm$^2$ is about 0.6~K. From the pressure dependence studies, this change is achieved at approximately 0.04 GPa. With a bulk modulus of 88 GPa (Ref. 45 in Ref.\cite{Ribeiro2022}), this corresponds to $\Delta V/V \approx 4.5 \times 10^{-4}$, which is more than 20 times larger than our largest estimate for the effect of irradiation. Therefore, steric effects are unlikely to be the reason for the disorder-induced enhancement of the transition temperature.

\section{Conclusion}

In conclusion, we observed an enhancement of the ferromagnetic transition temperature in LaCrGe$_3$ with the increasing concentration of point-like defects. This contradicts theoretical predictions of the suppression of itinerant ferromagnetism by disorder \cite{Belitz2005,Dresden}. We suggest that the flat band peak in $D(E)$ 0.25 eV below the $E_F$ broadens with disorder, causing an apparent increase in $D(E_F)$ which leads to an increase of $T_\text{FM}$ in this itinerant system. 

\section{Acknowledgements}

The research was supported by the U.S. Department of Energy, Office of Basic Energy Sciences, Division of Materials Sciences and Engineering.  Ames National Laboratory is operated for the U.S. Department of Energy by Iowa State University under Contract No.~DE-AC02-07CH11358. Electron irradiation was performed on SIRIUS platform supported by French National network of accelerators for irradiation and analysis of molecules and materials EMIR\&A under Project No. 21-3700.

\end{document}